\documentclass[10pt]{article}

\usepackage[superscript,sort]{cite}

\usepackage{ifthen,ifpdf}
\ifpdf
	\usepackage[pdftex]{hyperref}	\hypersetup{colorlinks=true,linkcolor=black,citecolor=black,urlcolor=blue,backref=page,bookmarks=true,breaklinks=true,plainpages=false }
	\usepackage[pdftex]{graphicx}
	\usepackage[usenames,pdftex]{color}
\else
	\usepackage[colorlinks=true,breaklinks=true,plainpages=false]{hyperref}
	\usepackage{graphicx}
	\usepackage[usenames]{color}
\fi

\usepackage{amsthm,amscd,amsxtra,amsfonts,amsmath,amssymb,multirow}
\usepackage{wrapfig}
\usepackage[footnotesize]{caption}
\usepackage[tiny,compact]{titlesec}
\usepackage[textwidth=0.8in,textsize=footnotesize]{todonotes}
\usepackage{algorithm,algorithmic,extarrows}

\begin{document}

\title{Multiresolution topological simplification
}

\author{
Kelin Xia$^1$ and
Zhixiong Zhao$^1$ and
Guo-Wei Wei$^{1,2,3}$ \footnote{ Address correspondences  to Guo-Wei Wei. E-mail:wei@math.msu.edu}\\
$^1$Department of Mathematics \\
Michigan State University, MI 48824, USA\\
$^2$Department of Electrical and Computer Engineering \\
Michigan State University, MI 48824, USA \\
$^3$Department of Biochemistry and Molecular Biology\\
Michigan State University, MI 48824, USA \\
}

\date{\today}
\maketitle

\begin{abstract}

Persistent homology has been devised as a promising tool for the topological simplification of complex data. However, it is computationally intractable for large data sets. In this work, we introduce  multiresolution persistent homology for tackling large data sets. Our basic idea is to match the resolution with the scale of interest so as to create a topological microscopy for the underlying data. We utilize flexibility-rigidity index (FRI) to access the topological connectivity of the data set and define a rigidity density for the filtration analysis. By appropriately tuning the resolution, we are able to focus the topological lens on a desirable scale. The proposed multiresolution topological analysis is validated by a hexagonal fractal image which has three distinct scales. We further demonstrate the proposed method for extracting topological fingerprints from DNA and RNA molecules. In particular, the topological persistence of a virus capsid  with 240 protein monomers is successfully analyzed which would otherwise be inaccessible to the normal point cloud method and unreliable by using coarse-grained multiscale persistent homology. The proposed method has also been successfully applied to the protein domain classification, which is the first time that persistent homology is used for practical protein domain analysis, to our knowledge. The proposed multiresolution topological method has potential applications in arbitrary data sets, such as social networks, biological networks and graphs. 

\end{abstract}

\newpage

{\setcounter{tocdepth}{5} \tableofcontents}

\newpage

\section{Introduction}

Proteins are of paramount importance to living organisms. They are essential to almost all the basic functions at the cellular level, such as providing structural support, regulating signal transduction, mediating gene transcription and translation, catalyzing metabolic reactions, etc.  It is commonly believed that protein functions are determined by protein  structures, the so called structure-function relationship. However, protein structures are, in turn, determined by the protein interactions. Protein interactions are  inherently of multiscale in nature, including short range covalent bonds, middle range hydrogen-bonds,  dipole-dipole interactions, van der Waals interactions, and long range electrostatic interactions. Consequently, protein structures are intrinsically multiscale as well, ranging from atomic scale, residue scale, alpha helix and beta sheet scale, domain scale in a single protein to protein scale in multiprotein complexes.  Geometric analysis of proteins is usually in terms of coordinates, bond length, bond angle, surface area, volume, curvature, etc, which often involve excessively high degrees of freedom and high dimensionality and can be computationally prohibitively expensive. Topological analysis of proteins is typically in terms of topological invariants, namely, connected components, tunnels or rings, and cavities or voids, which are zero dimensional (0D) and seldom useful. The  complexity and  multiscale nature of proteins or protein complexes call for innovative strategies in protein description, representation, characterization and analysis.

Recently, persistent homology has emerged as new approach for topological simplification of complex data \cite{Frosini:1999,Robins:1999, Edelsbrunner:2002,Zomorodian:2005}. The essential idea is to create a family of slightly different ``copies'' for a given data set through a filtration process so that the topology  of each copy can be analyzed. The copies are made different in the filtration process  either by the systematic increase in the radius of each sphere of a point cloud data or by the systematic change of the isovalue of a volumetric data. During the filtration process, the ``birth" and ``death" of topological invariants  i.e., Betti numbers, of the underlying copies can be tracked by using either   persistent diagrams or  barcode representation \cite{Ghrist:2008}. Appropriate mathematical apparatus has been deviced to organize simplicial complexes generated via the filtration process into homology groups \cite{Edelsbrunner:2002,Zomorodian:2005}. As such, persistent homology is able to provide a one-dimensional (1D) topological description of a given data set, in contrast to the  0D  description of the traditional topology and the high dimensional description of geometry. Therefore, persistent homology   introduces a geometric measurement to topological invariants, further bridging the gap between  geometry and topology.

Many efficient computational algorithms have been proposed for persistent homology analysis in the literature \cite{edelsbrunner:2010,Dey:2008,Dey:2013,Mischaikow:2013}. Persistent homology has found its success  in a variety of fields, including data analysis \cite{Carlsson:2009,Niyogi:2011,BeiWang:2011,Rieck:2012,XuLiu:2012}, image analysis \cite{Carlsson:2008,Pachauri:2011,Singh:2008,Bendich:2010,Frosini:2013},shape recognition \cite{DiFabio:2011}, chaotic dynamics verification \cite{Mischaikow:1999,Kaczynski:2004}, network structure \cite{Silva:2005,LeeH:2012,Horak:2009}, computer vision \cite{Singh:2008} and computational biology \cite{Kasson:2007,YaoY:2009, Gameiro:2013}.

However,  most successful applications of persistent homology are focused on topological characterization identification and analysis (CIA). Indeed, persistent homology has hardly been used for  physical modeling and/or quantitative prediction, to our knowledge. Recently, we have introduced persistent homology for mathematical modeling and prediction of  nano particles, proteins and other biomolecules \cite{KLXia:2014c, KLXia:2015a}. We have proposed  molecular topological fingerprint (MTF)   to reveal topology-function relationships in protein folding and protein flexibility \cite{KLXia:2014c}. We have employed  persistent homology to   predict the stability of proteins \cite{KLXia:2014c} and the curvature energies of fullerene isomers  \cite{KLXia:2015a,BaoWang:2014}. More recently, we have  proposed objective-oriented persistent homology  to proactively extract desirable topological   traits from complex data, based on variational principle \cite{BaoWang:2014}. Most recently, we have developed multidimensional persistent homology to achieve better characterization of biomolecular data \cite{KLXia:2014multi}. Persistent homology is found to provide an efficient approach for resolving ill-posed inverse problems in cryo-EM structure determination \cite{KLXia:2015b}.

\begin{figure}
\begin{center}
\begin{tabular}{c}
\includegraphics[width=0.95\textwidth]{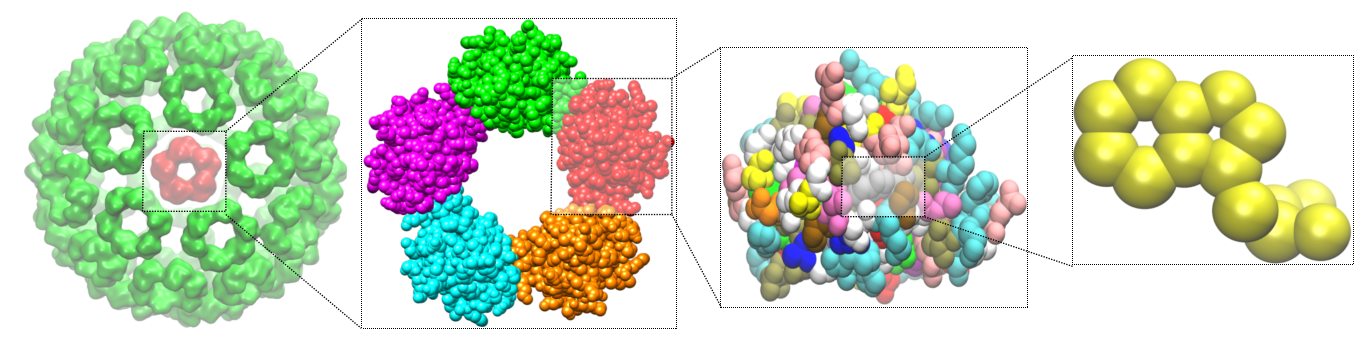}
\end{tabular}
\end{center}
\caption{Illustration of  multiscale features in a virus capsid structure (PDB ID: 1DYL). The capsid  has an icosahedral symmetry and consists of twelve pentagons and thirty hexagons. Each  hexagon or pentagon encompasses a multiprotein complex with five or six protein monomers. Each protein monomer has more than  a hundred amino acid residues. Each residue, in turn,  has many atoms.
}
\label{fig:Multiscale_illustration2}
\end{figure}

Figure \ref{fig:Multiscale_illustration2} illustrates the multiscale features of a virus particle. To understand the physical and biological properties of viruses and other macromolecular complexes, we need to have appropriate multiscale and multiresolution descriptions. The Protein Data  Bank (PDB) provides biomolecular structural information in a high level of detail, including atomic coordinates, observed sidechain rotamers, secondary structure assignments, as well as atomic connectivity. This type of structural data is usually known as point cloud data in persistent homology analysis. Multiscale description of biomolecules can be achieved in a variety of ways. Typically coarse-grained methods describe biomolecules in terms of superatoms or super-particles at a given scale. There are many superatom representations, including residue based, domain based and protein based ones. The corresponding  persistent homology analysis based on the residue representation has been explored in our earlier work \cite{KLXia:2014c}. Additionally, persistent homology analysis using protein based  coarse-grain representation  has been introduced in our recent work  for studying multiprotein complexes \cite{KLXia:2015b}. Nevertheless, coarse-grained persistent homology might suffer from  inconsistency due to the ambiguity in choosing the coarse-grained particle.

The direct application of persistent homology analysis to large biomolecules, such as the HIV virus capsid which has more than four million atoms, is unfeasible at present. One of reasons that lead to the failure is the use of a uniform resolution  in the filtration and cross-scale  filtration at a high resolution is prohibitively expensive in the present persistent homology algorithms.   Therefore, there is pressing need for innovative topological methods to deal with excessively large  data sets.

The objective of the present work is to introduce  multiresolution persistent homology (MPH). Our basic idea is to match the scale of interest with appropriate resolution in the topological analysis. In contrast to the original persistent homology that is based on a  uniform resolution of the point cloud data  over the filtration domain, the proposed MPH provides a mathematical microscopy of the topology at a given scale through a corresponding resolution. In spirit of wavelet multiresolution analysis, resolution based continuous coarse-grained representations are constructed for complex  data sets.    MPH can be employed to capture the topology of a given geometric scale and applied as a topological focus of lens. MPH becomes powerful when it is used in conjugation with  the data that has a multiscale nature.  For example,  one can use MPH to extract the topological fingerprints of a multiprotein complex  either at its atomic scale, residue scale, alpha helix and beta sheet scale, domain scale or at the protein scale.

The rest of this paper is organized as the follows. In  Section \ref{sec:theory}, we  introduce  multiresolution persistent homology. The underpinning multiresolution geometric modeling is accomplished by generalizing a flexibility rigidity index (FRI) method originally introduced for biomolecular data to general data. We design a hexagonal   fractal image to demonstrate the multiresolution analysis and investigate associated multiresolution topological persistence. In Section \ref{sec:biomolecule}, we explore multiresolution topological fingerprints of images and biomolecules. We reveal the  close relationship  between multiresolution geometry and multiresolution topology. We show that the FRI method provide a unified framework for both multiresolution geometric representation  and multiresolution topological analysis.  This paper ends with a conclusion.

\section{Method and algorithm}\label{sec:theory}

In this section, we introduce the theory and algorithm of  multiresolution geometric analysis and multiresolution persistent homology.  We construct the mutliresolution geometric representation by using the flexibility-rigidity index (FRI) method \cite{KLXia:2013d,Opron:2014}, which converts a point could data into a matrix or a density map. The conversion is modulated by a resolution parameter, which enables us to facilitate the multiresolution analysis of complex data. Additionally, FRI provides a resolution-controlled statistical average of a general data. The multiresolution topological analysis is developed based on the multiresolution representation of original data. To demonstrate the utility and examine validity of the proposed multiresolution geometric and topological methods, we design a hexagonal fractal image  with three distinct scales. We show that the proposed multiresolution persistent homology is able to extract the topological information at each of three scales. Therefore, the proposed topological method provides a topological microscopy of multiscale data at any desirable scale.

\subsection{Multiresolution geometric analysis}\label{sec:density}


Flexibility-rigidity index (FRI) \cite{KLXia:2013d,Opron:2014} was originally invented for the flexibility analysis of biomolecules. It provides an excellent prediction of macromolecular Debye-Waller factors or B-factors. The essential idea of FRI  is to construct flexibility index and rigidity index by certain kernel functions, and further use them to describe the topological connectivity of protein structures. In the present work, we generalize the FRI method for characterizing the rigidity and flexibility of arbitrary data sets, such as networks, graphs etc. The generalized FRI method facilitates the multiresolution geometric description of biomolecules, images and volumetric data in general.

 Assume that a data set  has a total $N$ entries, which can be  atoms, pixels or voxels   with generalized coordinates ${\bf r}_1, {\bf r}_2,\cdots, {\bf r}_N$.  In general, the rigidity index of the $i$th entry can be expressed as
\begin{eqnarray}\label{eq:rigidity}
\mu_i=\sum_{j}^N w_j\Phi(r_{ij};\eta_j),
\end{eqnarray}
where $r_{ij}=\|{\bf r}_i-{\bf r}_j \|$ is the generalized distance between the $i$th and $j$th entries, $w_j$ is a weight, which can be the element number of $j$th atom,   and  $\Phi(r_{ij};\eta_j)$ is a real-valued monotonically decreasing correlation function or probability density estimator \cite{GWei:2000} satisfying the following admissibility conditions
\begin{eqnarray}\label{eq:couple_matrix1-1}
\Phi( r_{ij};\eta_j)&=&1 \quad {\rm as }\quad  r_{ij}   \rightarrow 0\\ \label{eq:couple_matrix1-2}
\Phi( r_{ij};\eta_j)&=&0 \quad {\rm as }\quad  r_{ij}   \rightarrow\infty.
\end{eqnarray}
Here $\eta_j>0$ is a resolution parameter that can be adjusted to achieve the desirable resolution for a given scale.  Delta sequence kernels of the positive type discussed in an earlier work \cite{GWei:2000} are admissible correlation functions or kernels.  Commonly used  correlation functions are  generalized exponential  functions
\begin{eqnarray}\label{eq:couple_matrix1}
\Phi( r_{ij};\eta_j, \kappa) =    e^{-\left( r_{ij} /\eta_j \right)^\kappa},    \quad \kappa >0
\end{eqnarray}
or  generalized Lorentz functions
\begin{eqnarray}\label{eq:couple_matrix2}
 \Phi( r_{ij};\eta_j, \upsilon) =  \frac{1}{1+ \left( r_{ij} /\eta_j\right)^{\upsilon}},  \quad  \upsilon >0.
 \end{eqnarray}
Note that, in these functions, the larger the $\eta_j$ value, the lower the resolution is. The  flexibility index of the data set at the $i$th entry is defined as the inverse of the rigidity index
\begin{eqnarray}\label{eq:flexibility}
f_i=\frac{1}{\sum_{j}^N w_j\Phi(r_{ij};\eta_j)}.
\end{eqnarray}

Although the generalized distance $\|{\bf r}_i-{\bf r}_j \|$ can be regarded as the Euclidean space distance for biomolecular atoms, it is more generally defined in the present analysis, such as the distance between biological species or other entities. Therefore, the rigidity index and flexibility index are generalized concepts for arbitrary data sets, such as social networks, biological networks and graphs in the present formulation.

Flexibility index and rigidity index can be easily extended to  more general volumetric flexibility and rigidity functions. The rigidity function of the data can be expressed as,
\begin{eqnarray}\label{eq:rigidity_function}
\mu({\bf r})=\sum_{j}^N w_j\Phi(\parallel {\bf r}- {\bf r}_{j}\parallel;\eta_j)
\end{eqnarray}
and the flexibility function of the data can be given  as,
\begin{eqnarray}\label{eq:flexibility_function}
f({\bf r})=\frac{1}{\sum_{j}^N w_j\Phi(\parallel {\bf r}- {\bf r}_{j}\parallel;\eta_j)}.
\end{eqnarray}

The rigidity function can be regarded as the density distribution of a macromolecule, a picture or a volumetric data. Therefore, it provides an analytical representation of a data structure in $\mathbb{R}^3$. More importantly, the resolution  parameter in the rigidity function, i.e.,  $\eta_j$,  enables us to represent  the data at the scale of interest.  One  can set $\eta_j$ to a common constant $\eta$ for all atoms or data entries. If atoms in a biomolecular complex  are classified into subsets $\{\alpha_k\}$ according to residues, alpha helices, beta strands, domains or proteins,  one can also represent each subset of atoms  at a different resolutions $\eta_{\alpha_k}$. Similarly, one can choose $\eta_{\alpha_k}$ for subset $\eta_{\alpha_k}$ according other physical traits in general data.    Therefore, the resolution parameter $\eta_j$ offers two types of multiresolution representations: multiple common resolutions ($\eta$) and  a simultaneous multiresolution $\{ \eta_{\alpha_k}\}$.

Since  rigidity function gives rise to the density distribution of a macromolecule or a general data, the morphology of the macromolecule or data can be visualized at either a common resolution $\eta$ or a set of resolutions $\{ \eta_{\alpha_k}\}$. Obviously, the commonly used Gaussian surface \cite{ZYu:2008} is a special case of the present multiresolution geometric model. Additionally, the present multiresolution geometric model provides a basis for multiresolution persistent homology analysis.

\begin{figure}
\begin{center}
\begin{tabular}{c}
\includegraphics[width=0.9\textwidth]{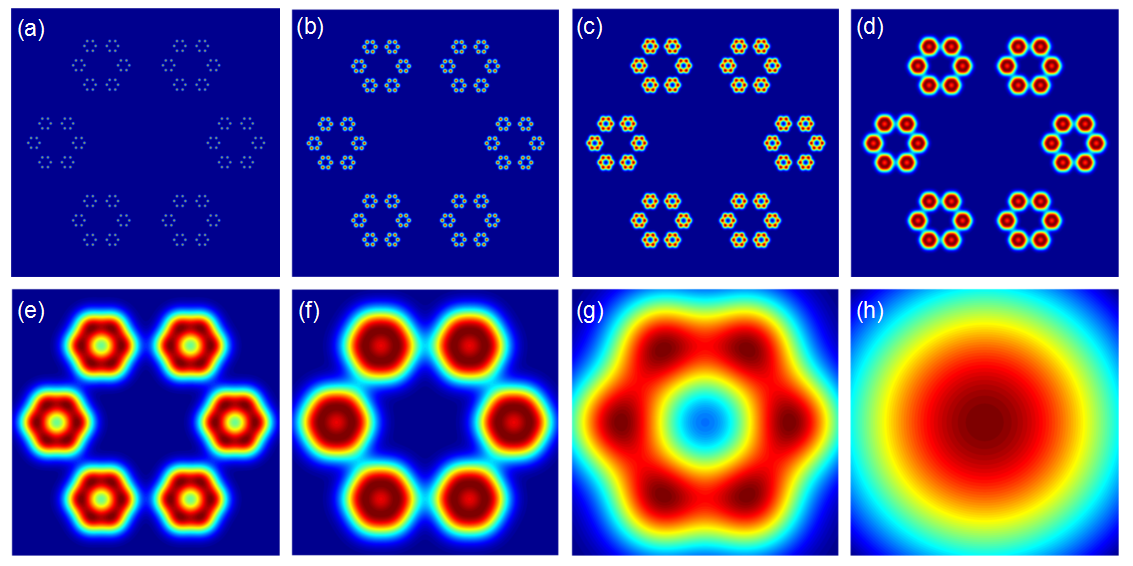}
\end{tabular}
\end{center}
\caption{Illustration of multiresolution geometric analysis of a 2D hexagonal fractal image. The rigidity functions $\mu({\bf r})$ are constructed from various resolution ($\eta$) values. From ({\bf a}) to ({\bf h}), $\eta$ values are set  to 0.2, 0.4, 0.6, 1.0, 3.0, 4.0, 10.0 and 30.0, respectively.
}
\label{fig:hexagon_density}
\end{figure}
Figure \ref{fig:hexagon_density} illustrates the multiresolution analysis of a hexagonal fractal image generated by rigidity function $\mu({\bf r})$ at various resolutions $\eta$.  It can be seen that rigidity functions give rise to a series of multiresolution geometric representations, focusing on different length scales of the original hexagonal fractal image. State differently, varying the resolution enables us to highlight the scale of interest.


\subsection{Multiresolution topological analysis}\label{sec:point_cloud}

To generate a persistent homology analysis,  a filtration process is used to construct a family of objects with different parametrizations. For a point cloud data,  the radius based filtration is most commonly employed. In this method, one associates each point with a ball whose radius is ever-increasing. When these balls gradually overlap with each other, simplicial complexes which are topological spaces are constructed by connecting corresponding points, line segments, triangles, and their high dimensional counterparts. There are various ways to decide which sets of points are to be connected at each radius. Among them, Vietoris-Rips complex (or Rips complex), in which a simplex is generated if the largest distance between any of its vertices is at most two times the ball radius, is most frequently used. As the radius increases, the previously formed simplicial complexes will be included into latter ones and a filtration process is thus created. In this manner, topological invariants that arises and perishes in this series of simplicial complexes are measured by their persistent time and visualized through a barcode representation. Sometimes, another representation called persistent Betti number (PBN) is used. Basically, PBN is the histogram of the total number of topological invariants over the filtration parameter.

The radius based filtration provides an efficient approach for the persistent homology analysis of point cloud data of relatively small data sets. However, it fails to capture two types of important physical properties of biomolecules. First, the information about element type, such as hydrogen (H),  carbon (C), oxygen (O), nitrogen (N), phosphorous (P), sulfur (S), etc,  is missing in the normal point cloud representation.  The difference in  elements has a dramatic impact to chemical and physical behaviors.  Additionally, the radius based filtration typically utilizes a uniform resolution of the point cloud data over the filtration domain, which is computationally expensive and requires huge memories for large biomolecules. Such  a description at a uniform resolution does not allow us to emphasize the topology at a given scale or study the topology with mixed scales for different parts.

We propose the MPH based on the  FRI method. Specifically, we construct  a matrix based on the FRI correlation function
\begin{eqnarray}\label{eq:filtrationM}
M_{ij}=\frac{w_j}{w_{\rm max}} (1-\Phi( r_{ij}; \eta_j)),
\end{eqnarray}
where $w_{\rm max}$ is the largest element number in the biomolecule or the largest weight in the data set, and $0\leq M_{ij}\leq 1$. Obviously, $M_{ij}$ can be viewed as the connectivity between the $i$th and $j$th atoms or entries. The smaller the $M_{ij}$ value is, the closer distance between the $i$th and $j$th entries is. A filtration over  matrix ($M_{ij}$) values can be constructed. Although a similar matrix based filtration method was introduced in our earlier work \cite{KLXia:2013d}, the multiresolution property based on the resolution parameter $\eta_j$ has never been explored. Compared with the radius based filtration, the FRI matrix based filtration incorporates appropriate resolution $\eta_j$ into the simplicial complex generation and can be used to highlight the topology at  a given scale of interest. In this work, we set $\eta_j=\eta$.  

Another multiresolution filtration can be constructed based on a series of isovalues of  the rigidity function volumetric data   ($\mu({\bf r})$)  shown in Eq. (\ref{eq:rigidity_function}). Unlike the commonly used  point cloud representation, rigidity function  incorporates  the information of atomic types and the resolution matching the desirable scale. In this manner, a multiresolution geometric representation can be obtained. By varying the resolution, our model can pinpoint to the local atomic detail or focus on global protein configuration. State differently, we can  focus the lens  on biomolecular traits of different scales, such as atom, residue, secondary-structure, domain, protein complex, organelle, etc. More importantly, the persistent homology analysis is employed in this multiresolution model to deliver a full ``spectrum" of topological characterization of the system. From the generated barcodes, a series of  topological fingerprints from various scales are obtained.

\subsection{Validation with a hexagonal fractal image}

\paragraph{Multiresolution geometric analysis of the hexagonal fractal image}

To demonstrate the proposed multiresolution geometric analysis and associated multiresolution topological analysis, we design a two-dimensional (2D) hexagonal fractal image as depicted in Figure \ref{fig:hexagon_density}. This structure is constructed by replacing each hexagonal vertex with a smaller hexagon. The coordinates of largest hexagon are set to ($\sqrt{30}$, 10), ($\sqrt{30}$, -10), (0.0, 20), (0.0, -20), (-$\sqrt{30}$, 10), and (-$\sqrt{30}$, -10). It is easy to seen that the length of the edge is 20. These nodes are then used as centers for second-level hexagonal structures. The edge length of these hexagons is 0.25 times of the original edge length (i.e., 5). By removing all the nodes of the original hexagon, we have a totally 36 new nodes and 6 smaller hexagons in our second level structure. By repeating this process again, i.e., replacing each node in the second-level hexagons with a third-level hexagonal structure,  setting the edge length to be 0.25 times of the second level edge length (i.e., 1.25), and removing all the second level nodes, we arrive at our final  hexagonal fractal image with 216 nodes in 36 third-level hexagons.

The FRI based multiresolution analysis  of the hexagonal fractal image is depicted in  Figure \ref{fig:hexagon_density}. Obviously, there are three scales in the fractal image. The smallest scale is better resolved at a resolution around 0.4. The middle scale, which shows six hexagons,  is reflected at resolution of 1.0-3.0. At the large scale, there is only one hexagon which is better represented at the resolution of  4.0.

\begin{figure}
\begin{center}
\begin{tabular}{c}
\includegraphics[width=0.5\textwidth]{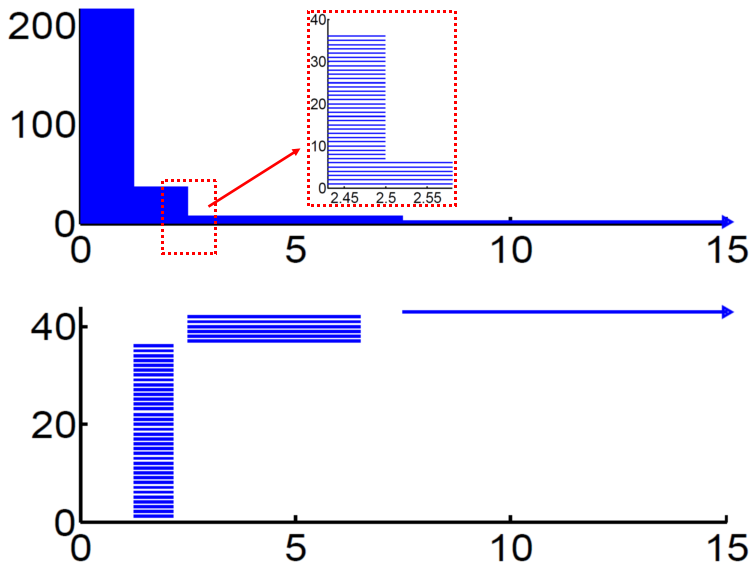}
\end{tabular}
\end{center}
\caption{ Topological fingerprints of a hexagonal fractal image.
Top and bottom panels are for $\beta_0$ and $\beta_1$ barcodes, respectively. The
horizontal axis is the filtration parameter, i.e., radius.  Note the separation of ring structures in the $\beta_1$ barcode.
}
\label{fig:hexagon_pattern}
\end{figure}

\paragraph{Topological fingerprint of the hexagonal fractal image}

Typically,  barcodes are obtained from persistent homology analysis. In our study, we arrange the barcodes in a sequence according to their birth time. The resulting barcode pattern for a given data set is called a topological fingerprint, which can be used to  identify the data set. The topological fingerprint of the hexagonal fractal image   can be obtained from persistent homology analysis. Figure \ref{fig:hexagon_pattern} demonstrates the persistent barcodes of  the hexagonal fractal  generated by the radius filtration. The upper and lower panels are barcode representations of $\beta_0$ and $\beta_1$, respectively. For all our persistent barcodes, the horizontal axis is filtration parameter, i.e., either the radius (\AA)  or the rescaled density value.

\begin{figure}
\begin{center}
\begin{tabular}{c}
\includegraphics[width=0.9\textwidth]{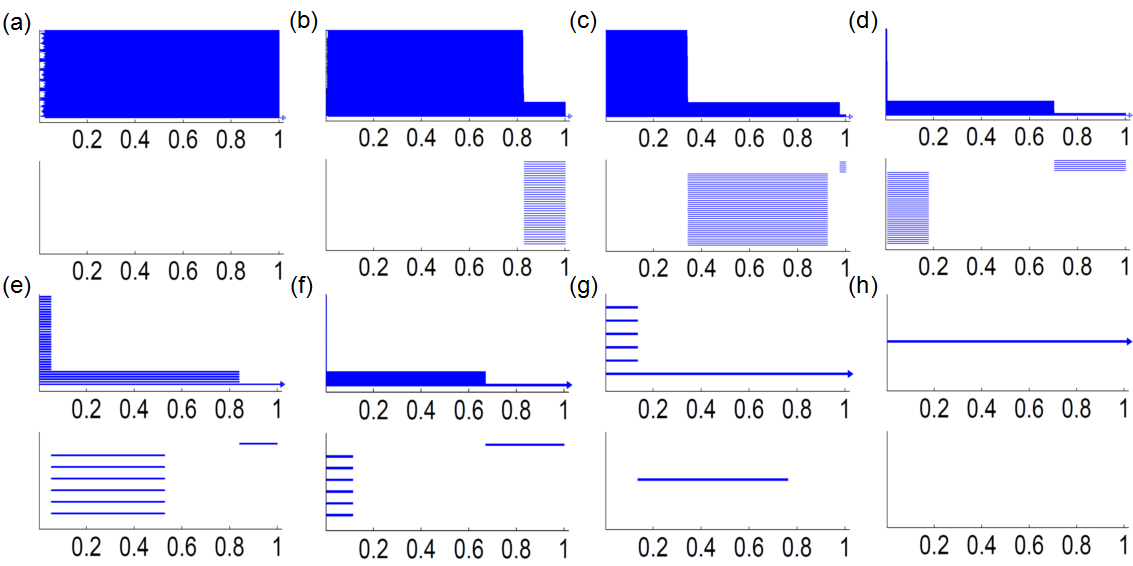}
\end{tabular}
\end{center}
\caption{  Multiresolution persistence of the hexagonal  fractal image. Top and bottom panels are for $\beta_0$ and $\beta_1$ barcodes, respectively.  The horizontal axises denote the rescaled rigidity density value. The topological fingerprints at different scales can be identified.
From ({\bf a}) to ({\bf d}),   persistent barcodes are generated at resolutions $\eta= 0.2, 0.4, 0.6, 1.0, 3.0, 4.0, 10.0$ and 30.0.
}
\label{fig:haxagon_barcodes}
\end{figure}

Topologically, $\beta_0$ is for isolated components and $\beta_1$ represents one-dimensional loops or rings. It is seen that originally there are 216 $\beta_0$ bars, corresponding to  216 nodes in the hexagonal fractal. When filtration size goes to 1.25, most of these $\beta_0$ bars are simultaneously killed and the total number of isolated components reduces to 36. This means that 1-simplexes (edges) begin to form between adjacent 0-simplexes (nodes) in the related Vietoris-Rips complex, eliminating isolated components. Meanwhile, 36 individual $\beta_1$ bars, i.e., 36 hexagonal rings, emerge simultaneously. With the advance of the filtration, the persistent Betti numbers (PBNs) of $\beta_0$ bars undergo two further reductions. First,  PBN descends to 6 at 2.5
and  to 1 at 7.5.  While two types of larger scaled $\beta_1$ bars have been generated. From a topological point of view, the detailed small-scale structures are removed by  the creation of 2-simplexes, at the same time higher level structures appear when more connections (1-simplexes) are established. In general, it is seen that all three levels of hexagonal structures are captured in both $\beta_0$ and $\beta_1$ bars. All of these identical bars form a unique topological pattern which is directly related to their structure properties, and thus is called a topological fingerprint for the fractal image.

\begin{figure}
\begin{center}
\begin{tabular}{c}
\includegraphics[width=0.9\textwidth]{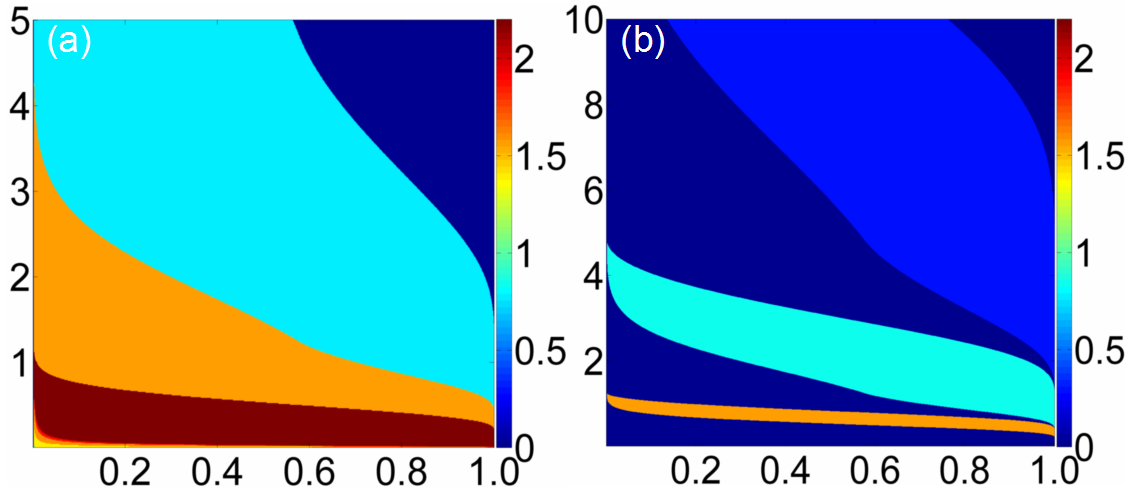}
\end{tabular}
\end{center}
\caption{Illustration of 2D persistent homology in terms of PBNs for the hexagonal fractal image. The horizontal axises denote the rescaled rigidity density value, and the vertical axises represent the resolution $\eta$. The logarithm of  PBNs for $\beta_0$  and $\beta_1$ are plotted in (a) and (b), respectively. Three large bands in both  $\beta_0$  and $\beta_1$ indicate three distinguished  scales of the fractal topology.
}
\label{fig:hexagon_multiscale}
\end{figure}

\paragraph{Multiresolution topological analysis of the hexagonal fractal image}

To generate a multiresolution topological analysis of the multiscale hexagonal fractal image, we use the exponential function with power $\kappa=2$ and systematically changes the resolution $\eta$. The computational domain $\Omega$ is set as $\Omega=[-30,30]\times [-30,30]$, and the grid spacing is chosen 0.05 in order to capture all the local details in the structure.  To avoid confusion, we linearly rescale all the rigidity function value to the region [0,1] using equation
\begin{eqnarray}\label{eq:rescale}
\mu({\bf r})^{s}=1-\frac{\mu(\bf r)}{\mu_{\rm max}}, ~\quad~ \forall {\bf r} \in \Omega,
\end{eqnarray}
where  $\mu({\bf r})$ and  $\mu({\bf r})^{s}$  are the original and rescaled rigidity density value, respectively. Here $\mu_{\rm max}$ is the largest density value in the original data. The rescaled density value is then used as the filtration parameter. Figure \ref{fig:haxagon_barcodes} depicts the multiresolution topological analysis of the hexagonal fractal image. Clearly, at the resolution of 0.2, the topology is a set of 216 isolated nodes. At the resolution of 0.4, the topology shows the formation of 36 small hexagons from 216 nodes. At the resolution of 3.0, each small-scale hexagon becomes a ``superatom''.  Only the middle-scale topological features appear, namely  the formation of 6 middle-scale hexagons from 36 ``superatoms''.    One can also see the formation of a large ring from 6 middle-scale hexagons. At the resolution of 10.0, No detailed topological structure of the six middle scale  hexagons is visible. The topology shows the formation of the large-scale ring from 6 ``superatoms''. At the lowest resolution of 30.0, all of the 216 nodes are topologically equivalent to a superdot and there is no ring structure at all.          It is seen that the varying of the resolution gives rise to  a series of topological representations of the original structure at various scales.  State differently, the proposed multiresolution persistent homology provides an adjustable ``topological lens", and enables us to focus on any scale of interest.

The  multiscale nature of the resolution parameter can be more clearly seen in a multi-dimensional filtration process. Figure \ref{fig:hexagon_multiscale} demonstrates the PBNs under various resolution values. The horizontal axises represent the rescaled rigidity density value, and the vertical axises are the resolution. The color bar indicates the common logarithm (logarithm with base 10) of the PBN  values. It should be noticed that in order to avoid the situation of $log_{10}(0)$, we systematically increase all the PBNs by one in both $\beta_0$ and $\beta_1$ plots. As demonstrated in Figure \ref{fig:hexagon_multiscale}, there exist three   bands in both $\beta_0$ and $\beta_1$ plots. Each band represents a scale in the structure, capturing the essential multiscale topological properties of the hexagonal fractal image.

\section{Multiscale multiresolution  topological analysis of biomolecules}\label{sec:biomolecule}


In this section, we illustrate how to employ the proposed multiresolution topological analysis for the study of biomolecular data. We first use a DNA structure to demonstrate multiscale geometric analysis generated at two atomic scale representations and one coarse-grained (nucleic acid) representation in terms of phosphorus atoms. The associated topological analyses provide topological fingerprints for biomolecules at different scales.
We further illustrate multiresolution geometric and topological analysis of biomolecular data. Specifically, we construct biolmolecules at various FRI resolutions using the rigidity function in our FRI method. The rigidity density, i.e., volumetric profile of the  rigidity function, is used to create multiresolution topological  analysis by varying the resolution $\eta$. Finally, we apply the present multiresolution persistent  homology to protein domain classification. We show that at appropriate resolutions, the $\beta_0$ invariants indicate the separation of domains.

\subsection{Multiscale   topological analysis of biomolecules}

As discussed earlier, macromolecules are intrinsically multiscale. On the one hand, persistent homology analysis of macromolecular data gives rise to multiscale persistence. On the other hand, the multiscale nature of macromolecules allows us to carry out the persistent homology analysis at different scales. For example, multiprotein complexes can be represented at a variety of scales, including atomic scale, residue scale, domain scale and protein scale. The different representations  of a multiprotein complex lead to different topological fingerprints and associated topological interpretations.

\begin{figure}
\begin{center}
\begin{tabular}{c}
\includegraphics[width=0.6\textwidth]{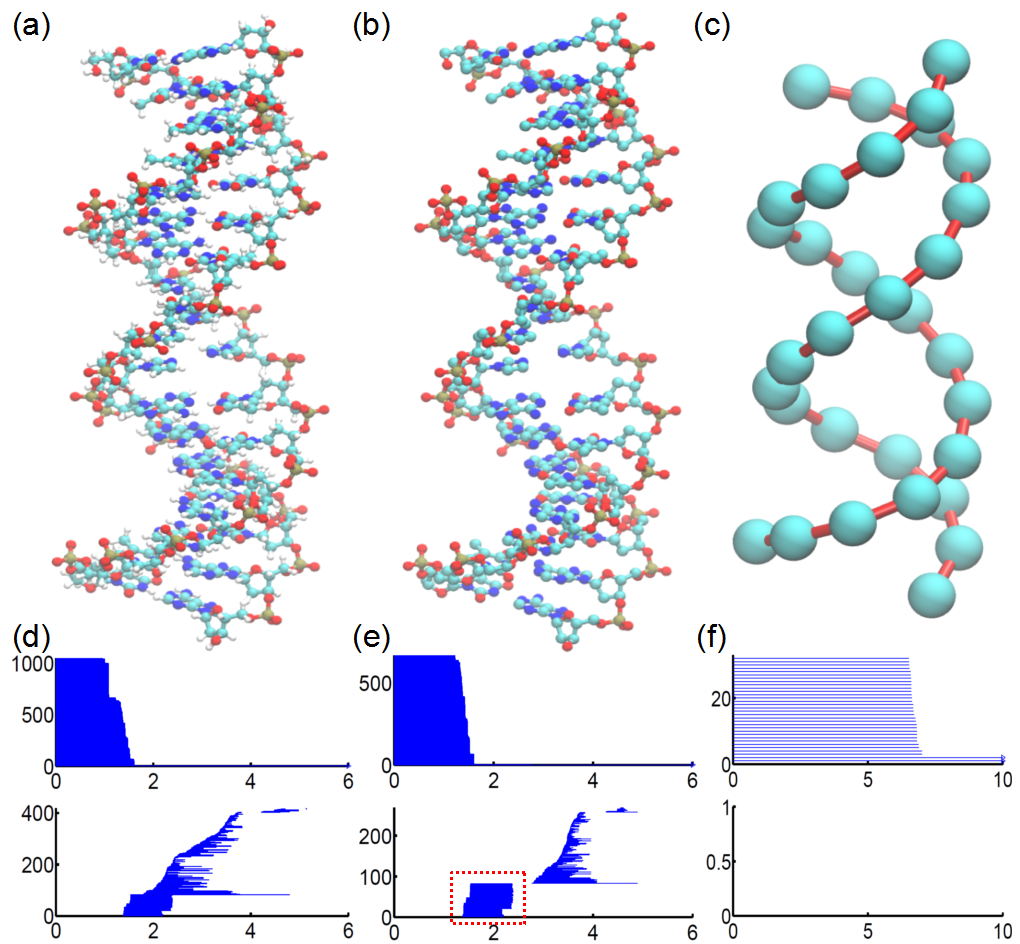}
\end{tabular}
\end{center}
\caption{Multiscale analysis of DNA molecule (PDB ID: 2M54).
({\bf a})-({\bf c}) Multiscale geometric analysis of the DNA molecule in all-atom representation, all-atom representation without hydrogen atoms, and coarse-grained representation (using phosphorous  atoms), respectively; ({\bf d})-({\bf f}) Corresponding persistent barcodes for the above three representations. Top and bottom panels are for $\beta_0$ and $\beta_1$ barcodes, respectively. The horizontal axises denote the filtration parameter, i.e., radius (\AA).  In all-atom representation, nearly half of the $\beta_0$ bars end around 1.2\AA~ in ({\bf d}). With the removal of the hydrogen atoms, local topological invariants representing hexagonal and pentagonal rings are clearly separated from the global topological structures  in ({\bf e}). The dimensionality reduction with the coarse-grained model displays not only the signature of phosphorous atoms, but also the double string structure   in ({\bf f}).
}
\label{fig:dna_pattern}
\end{figure}

Figure \ref{fig:dna_pattern} illustrates a multiscale persistent homology analysis of a DNA molecule (PDB ID: 2M54). The DNA molecule is described in  all-atom representation (Figure \ref{fig:dna_pattern}(a)), all-atom representation without hydrogen atoms (Figure \ref{fig:dna_pattern}(b)), and  coarse-grained phosphorous representation (Figure \ref{fig:dna_pattern}(c)). The corresponding topological fingerprints shown in Figures \ref{fig:dna_pattern}(e)-(f) are dramatically different. The building blocks of nucleotides include nitrogenous base, five-carbon sugar, and phosphate group. The five-carbon sugar has a pentagonal ring (PR) and the nitrogenous base has either a hexagonal ring (HR) or one HR and one PR. It can be seen that these local structural details are well-resolved in the topological fingerprint computed from our hydrogen-free all-atom data as indicated in Figure \ref{fig:dna_pattern}(e).   The signatures of HR and PR appear around 2.0 \AA, which is very similar to  PRs and HRs in protein molecules \cite{KLXia:2014c}, indicating they are consistent in all biomolecular structures. The  selection of coarse-grained  DNA models is very subtle.  In our case, a phosphorous atom is chosen  to generate a representation of a nucleotide. In this manner, only the two strand backbones are preserved, which is represented by two long persisting bars in $\beta_0$. There is no ring formation or $\beta_1$ bar in the coarse-grained topology.

For multiprotein complexes, such as viruses and microtubules, it is computationally too expensive to directly compute the full spectrum of topological fingerprints in the atomic scale. Multiscale persistent homology based coarse-grained residue or protein representations provide a potential solution to this problem.  However, one always faces a difficulty as to how to select representative superatoms for a given atomistic data. Different choices lead to dramatically different topological fingerprints.  Large topological errors can be generated by using undesirable superatoms. These issues highlight the problematic nature of multiscale persistent homology method.

The multiresolution persistent homology proposed in this work by-passes such a difficulty. It effectively provides a coarse-grained representation at a large scale. However, unlike the aforementioned superatom-based coarse-grained representation, the present resolution based coarse-grained method provides a faithful representation of the original geometry.

\subsection{Multiresolution topological analysis of biomolecules}

It is well known that in biomolecules, there are various types of atoms,  H, C, O, N, P, S, etc., with a wide range of element numbers. As discussed in Section \ref{sec:point_cloud}, traditional point cloud representation does not discriminate these chemical elements and  treats their equally.
Our FRI based density model can automatically take the element information into consideration. More importantly, the resolution can be controlled to match the scale of interest.  This enables us to tackle the topology of macromolecules at large scales that are intractable with the current point cloud approaches. In this section, we apply multiresolution persistent homology to  two biomolecular systems, i.e., a complex RNA structure and a virus capsid structure.

\begin{figure}
\begin{center}
\begin{tabular}{c}
\includegraphics[width=0.5\textwidth]{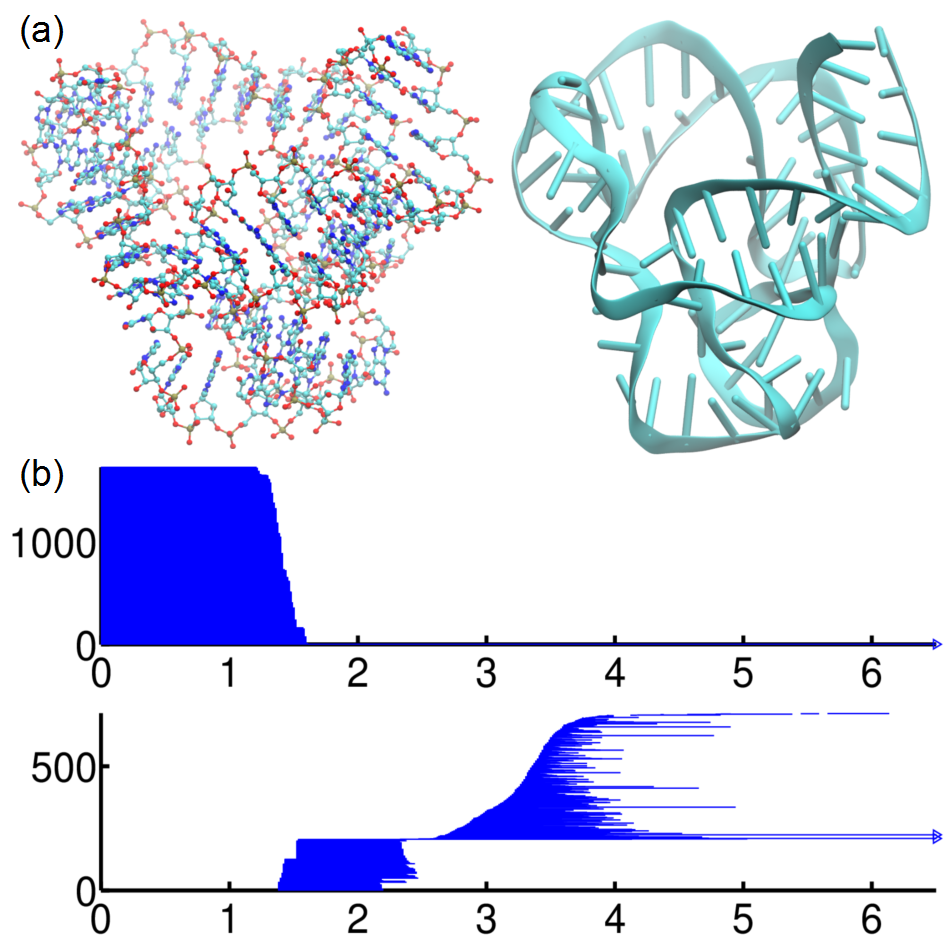}
\end{tabular}
\end{center}
\caption{ Geometric and topological analysis of an RNA molecule (PDB ID: 4QG3).
({\bf a}) The atom-bond representation and cartoon representation of RNA 4QG3. The atom-bond representation reveals more local nucleotide information. The cartoon representation focuses more on global RNA helix configuration.
({\bf b}) The persistent barcodes for RNA 4QG3. Top and bottom panels are for $\beta_0$ and $\beta_1$ barcodes, respectively. The horizontal axis denotes the filtration parameter, i.e.,  radius (\AA).
As there is no hydrogen atoms in the data, local pentagon and hexagon rings from nucleotide base and sugar, and global helix strips  appear in two different domains in the $\beta_1$ panel.
}
\label{fig:rna_PointCloud}
\end{figure}

\paragraph{A complex RNA structure}

A complex RNA molecule is used to demonstrate the utility of the present multiresolution topology analysis for biomolecules.  To prepare the data, we remove the protein and all ions in biomolecule 4QG3, and retain only the RNA part. Figure \ref{fig:rna_PointCloud} depicts the basic structure and its persistent barcodes generated from point cloud data. The pentagon and hexagon rings are represented by the short bars that lasts from around 1.3 to 2.5\AA~ in the $\beta_1$ panel. When radius filtration goes further, global topological invariants appears. Among them, there are two long-persistent $\beta_1$ bars, which persist after even our filtration size 7.0\AA. These two bars reveal the intrinsic topological invariants in the RNA structure, i.e., two large loops formed by the backbone of the RNA molecule.
\begin{figure}
\begin{center}
\begin{tabular}{c}
\includegraphics[width=0.5\textwidth]{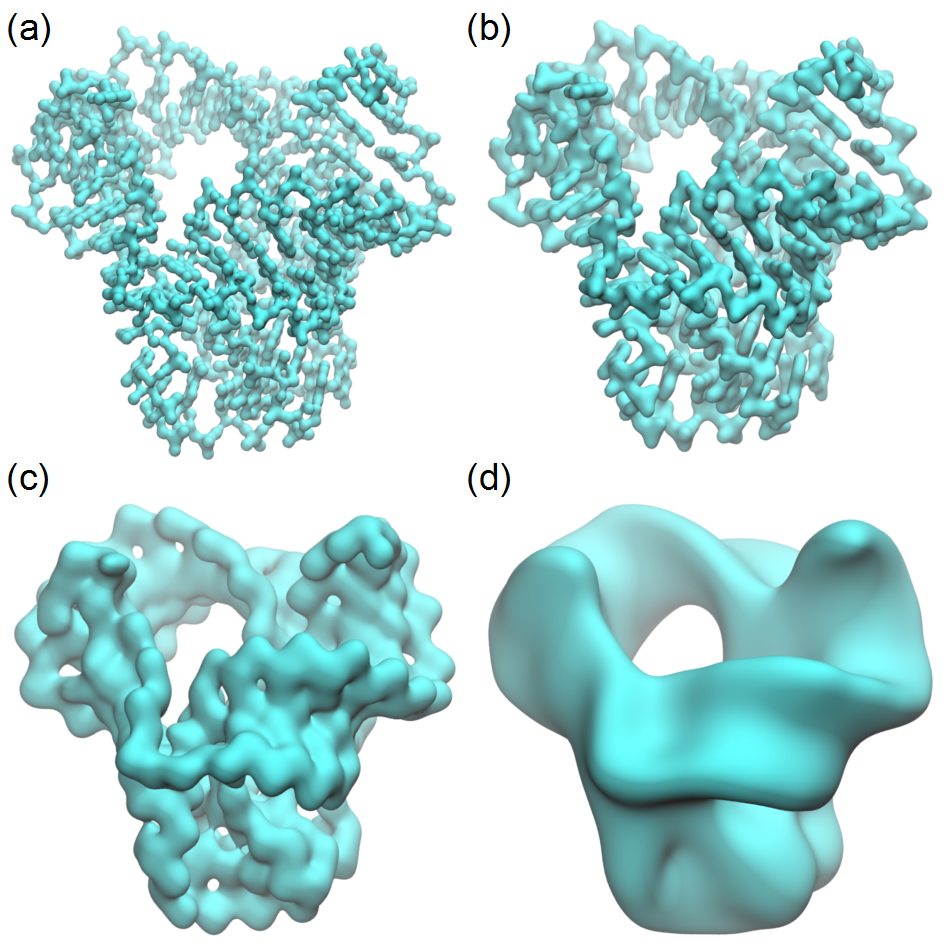}
\end{tabular}
\end{center}
\caption{ Multiresolution geometric analysis of RNA 4QG3. At various resolutions, rigidity density profiles emphasizing on different scales of RNA 4QG3 structure. ({\bf a})-({\bf d}) are RNA surfaces extracted from density profiles generated with resolutions $\eta=0.7, 1.0, 2.0$ and  4.0\AA, respectively. It can be seen  from ({\bf a}) that, the rigidity density map focuses on the atom-and-bond scale. The pentagon and hexagon rings in the base and sugar part are well captured. More global information begins to reveal when the resolution parameter increases in ({\bf b}). The RNA double helix string pattern is visible in ({\bf c}). The minor grove and major grove can be identified and the loops formed by the helix string are revealed. Further increase in resolution value smears most of the local information, leading to only the intrinsic loop.   }
\label{fig:rna_surface_multiscale}
\end{figure}

Next, we illustrate the multiresolution analysis of RNA 4QG3. First, we take into consideration of
atom types  by setting $w_j$ in the FRI correlation function to their element numbers. Additionally,  we vary the FRI resolution  $\eta$ from 0.3 to 4.0\AA~ to deliver a full ``spectrum" of geometric resolution in our rigidity density map.     In our persistent homology analysis, we still linearly rescale all the density maps to the range of [0,1] using Eq. (\ref{eq:rescale}). The RNA molecule extracted from RNA-protein complex 4QG3 has 1723 atoms and large loops in its structure. Since small grid spacing can be prohibitively expensive for this system,  we use a grid spacing of 0.3 \AA~ in our study. As a result, some detailed local topological structures may not be fully resolved and may even appear as noise in our persistent barcodes. Therefore, in our barcode results, we removes all the bars with persistent length less than 0.05 with respect to a total length of 1.

\begin{figure}
\begin{center}
\begin{tabular}{c}
\includegraphics[width=0.9\textwidth]{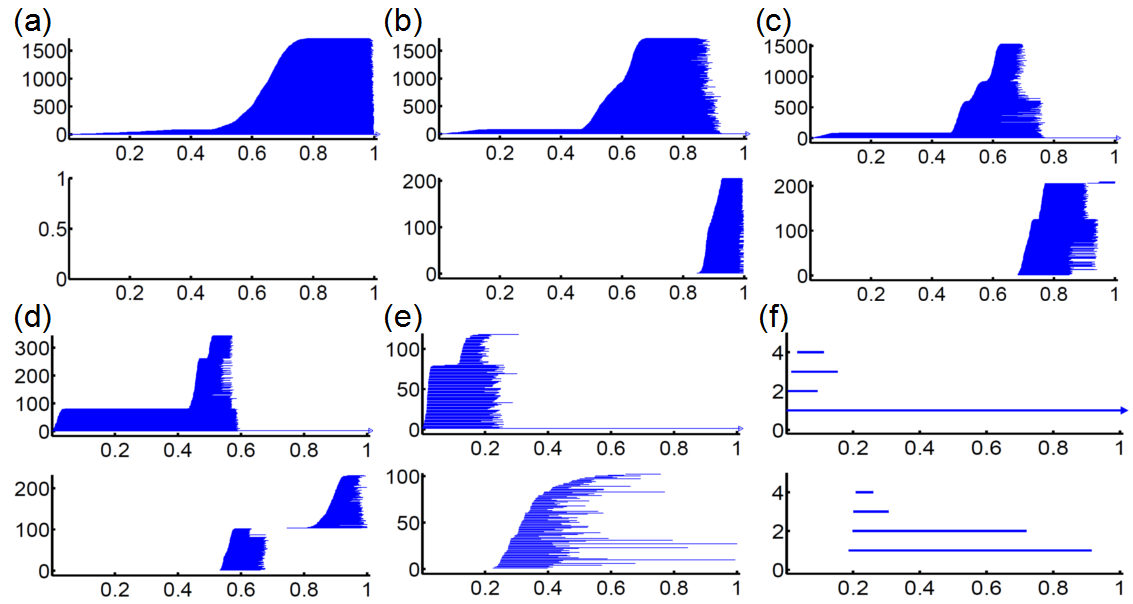}
\end{tabular}
\end{center}
\caption{ Multiresolution  topological analysis of   RNA molecule extracted from RNA-protein complex 4QG3.
({\bf a})-({\bf f})  Persistent barcodes for RNA 4QG3 density profiles generated at resolutions  $\eta=0.3, 0.5, 0.7, 1.0, 2.0$ and  4.0\AA, respectively. Top and bottom panels are for $\beta_0$ and $\beta_1$ barcodes, respectively. The horizontal axises denote the rescaled rigidity density value. It can be seen that at various resolutions, the persistent barcodes give a clearly demonstration of various scales existed in the structure. In ({\bf a}), only the atomic information can be seen.
Local pentagon and hexagon ring structure appears in ({\bf b}). Global topological invariants emerge in ({\bf c}) and gradually become  dominant in ({\bf d}) and ({\bf e}). Further increase in the resolution value  eliminates most transitional local topological invariants, leaving two largest intrinsic loops as demonstrated in ({\bf f}). It is obvious that when the resolution parameter reaches a certain limit, all topological invariants will be gone and the density map of the whole RNA molecular will melt into a feature-less body  as seen in the hexagonal fractal image.
}
\label{fig:rna_multiscale}
\end{figure}

The rigidity density maps generated by various resolutions have dramatically different physical implications. The isosurfaces extracted from these maps give a good explanation of the present multiresolution analysis. Figure \ref{fig:rna_surface_multiscale} demonstrates four isosurfaces from density data generated by $\eta= 0.7, 1.0, 2.0 $ and  4.0\AA, respectively. It can be seen that with the increase of $\eta$ value, isosurfaces gradually shift from a  local type of scales to a   global type of scales. More specifically, when $\eta$ is smaller than  0.7\AA, generated density maps focus on the scale of atom and atom-bond. When $\eta$ is increased to around 1.0\AA, nitrogenous base or five-carbon sugar scale dominates. The further increase of $\eta$ to around 2.0\AA ~leads to the major groove and minor groove scale.  Finally, when $\eta$ goes beyond 4.0\AA, rigidity map of the RNA  gradually ``melt" into a single gigantic body. This resolution shifting generates the corresponding topological changes as can be clearly observed from our persistent barcodes.  In our multiresolution persistent homology analysis, we systematically change $\eta$ from 0.3\AA~ to 4.0\AA.  As demonstrated in Figure \ref{fig:rna_multiscale},   total PBNs in $\beta_0$ panels,  gradually decrease from 1723 to 4 and will finally dwindle into 1 if we increase $\eta$ value further. This phenomenon indicates the inverse relationship between the topological complexity and the resolution value. Additionally,  there are 79 $\beta_0$ bars that appear much more earlier in the filtration process. These bars are due to 79 phosphorous atoms in the RNA structure, as they have a much larger element number.  For $\beta_1$ bars, more intriguing patterns can be observed. Originally there are 205 $\beta_1$ bars, i.e., the total number of  PRs and HRs in local nitrogenous bases and five-carbon sugar rings. The number of $\beta_1$ bars soars up when more global topological invariants are captured. However, the further increase in  the resolution parameter results in the loss of local topological invariants, and thus the PBNs  gradually decline. By  increasing the resolution parameter, we are able to identify more intrinsic global topological properties in the structure.

\paragraph{A Virus capsid structure}

The virus capsid structure of 1DYL shown in Figure \ref{fig:1dyl_multiscale} has multiple scales, ranging from atomic, residual, protein scales to pentagonal or hexagonal protein complex scales. There are 5705 atoms in each of 12 pentagon-shaped complexes and 6844 atoms in each of 30 hexagon-shaped complexes, leading to a total of 273780 atoms in capsid. Computationally, it is prohibitively expensive to incorporate all the scales in a uniform topological representation. Therefore, we first analyze  the building block, i.e., pentagon-shaped and hexagon-shaped protein complexes.

\begin{figure}
\begin{center}
\begin{tabular}{c}
\includegraphics[width=0.8\textwidth]{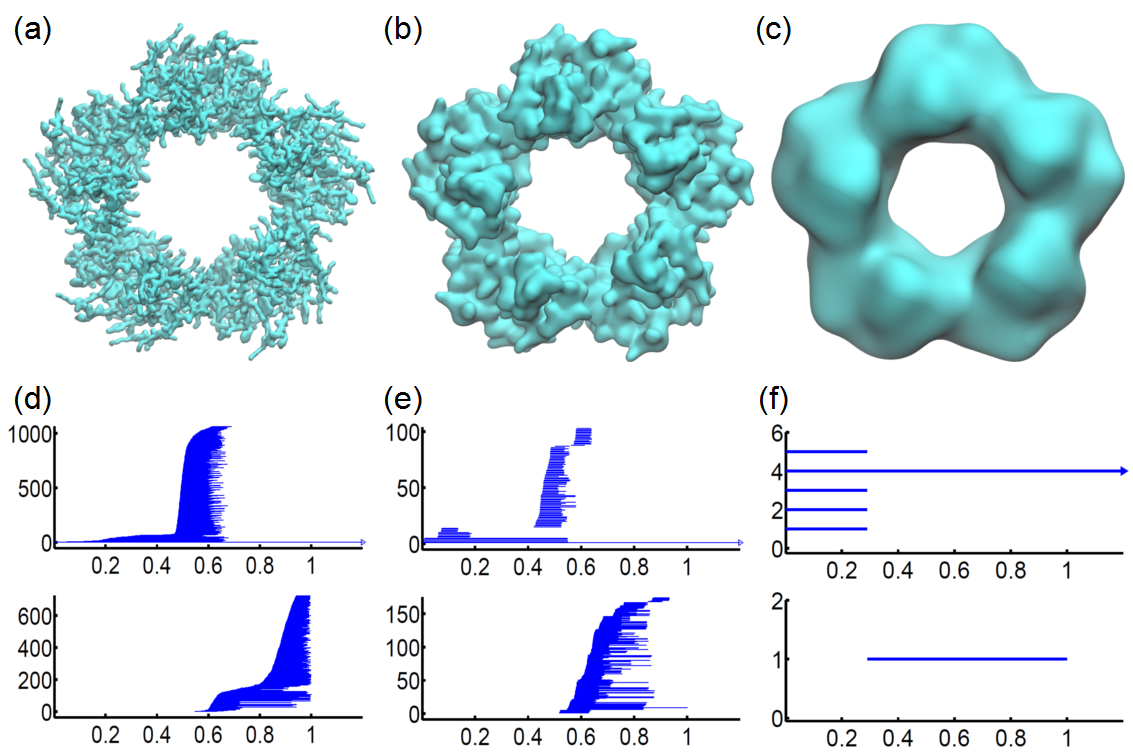}
\end{tabular}
\end{center}
\caption{ Multiresolution   analysis of  pentagon-shaped  protein complex constructed from 1DYL.
({\bf a})-({\bf c}) Isosurfaces of rigidity density profiles generated at three resolutions   $\eta= 1.0, 2.0$ and  6.0\AA;
({\bf d})-({\bf f}) The corresponding persistent barcode representations. Top and bottom panels are for $\beta_0$ and $\beta_1$ barcodes, respectively.  The horizontal axises denote the rescaled rigidity density value. It can be seen that with the decrease of resolution,  the density profiles of protein isosurfaces shift from local details to global patterns. The persistent barcodes become simpler as $\eta$ increases.   It worth mentioning  that the pentagon structure is highly symmetric as can been seen from the five identical $\beta_0$ bars in ({\bf f}).
}
\label{fig:PC_pentagon_pattern}
\end{figure}

In the construction of  rigidity density maps, we incorporate atom type information by the association of weight parameters $w_j$ with the atomic element number.  For such a large system, we use grid size 0.6 \AA~ in our density map constructions for a single pentagon-shaped complex or a single  hexagon-shaped protein complex. However, the constructed density map for the virus capsid is much larger in size. We therefore set the grid size to 2.0\AA~. We also linearly rescale all the generated density maps to the range of [0,1] using Eq. (\ref{eq:rescale}) and remove the bars with length less than 0.05 in  barcodes. Note that PDB provides the structure information for a single protein and related symmetry operations. Therefore, pentagon-shaped protein complexes and pentagon-shaped protein complexes, as well as the virus capsid, are all constructed using the symmetry information in the PDB for protein 1DYL.

\begin{figure}
\begin{center}
\begin{tabular}{c}
\includegraphics[width=0.8\textwidth]{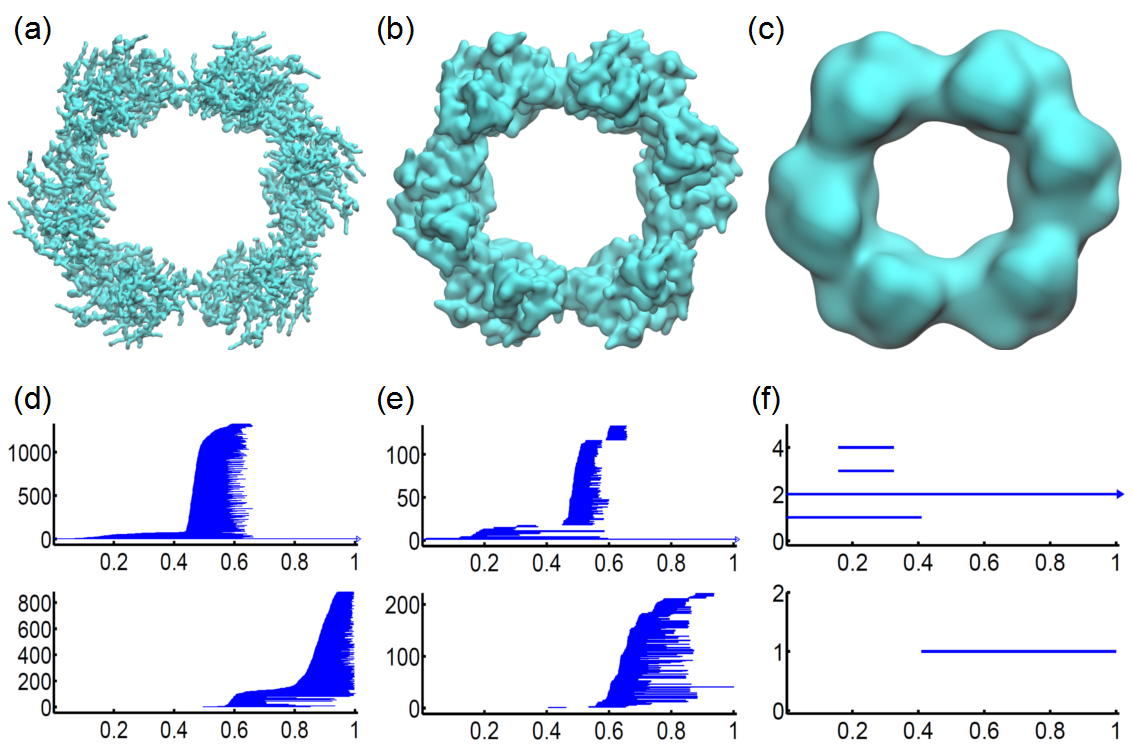}
\end{tabular}
\end{center}
\caption{ Multiresolution   analysis of of hexagon-shaped protein  complex constructed from 1DYL.
({\bf a})-({\bf c}) Isosurfaces of rigidity density profiles generated at three resolutions   $\eta= 1.0, 2.0$ and  6.0\AA;
({\bf d})-({\bf f}) The corresponding persistent barcode representations. Top and bottom panels are for $\beta_0$ and $\beta_1$ barcodes, respectively.  The horizontal axises denote the rescaled rigidity density value. It can be seen that similar to the case of the pentagon-shaped protein complex, the atomic details in the density profiles disappear as the resolution decreases. This scenario is also confirmed by the dramatically reduction in the numbers of persistent bars in both $\beta_0$ and $\beta_1$ panels when resolution decreases (i.e., $\eta$ increases). However, one should note that the hexagon-shaped protein complex has a unique asymmetric pattern. Instead of six identical $\beta_0$ bars, there are only four  $\beta_0$ bars existed in ({\bf f}), two appearing earlier and two emerging later during the filtration.
}
\label{fig:PC_hexagon_pattern}
\end{figure}

To analyze the topological properties of pentagon-shaped and hexagon-shaped protein  complex, we generate the density data at resolutions  $\eta= 1.0, 2.0$ and  6.0\AA. It should be noticed that if the resolution value is smaller than  the grid size 0.6\AA,  local structures cannot be well captured and the resulting  barcodes  tend to be very noisy. For  pentagon-shaped protein complex,  Figures  \ref{fig:PC_pentagon_pattern}({\bf a})-({\bf c}) demonstrate the representative isosurfaces extracted from different $\eta$ values. It can be seen that structure details gradually blur and merge to a pentagon-shaped body, indicating the scale changes.  Persistent barocode results are illustrated in Figures \ref{fig:PC_pentagon_pattern}({\bf d})-({\bf f}). The PBNs for $\beta_0$ and $\beta_1$ consistently dwindle. The five individual proteins and the pentagonal loop structure are well captured in $\beta_0$ and $\beta_1$ panels in $\eta= 6.0$ \AA~ cases.

\begin{figure}
\begin{center}
\begin{tabular}{c}
\includegraphics[width=0.9\textwidth]{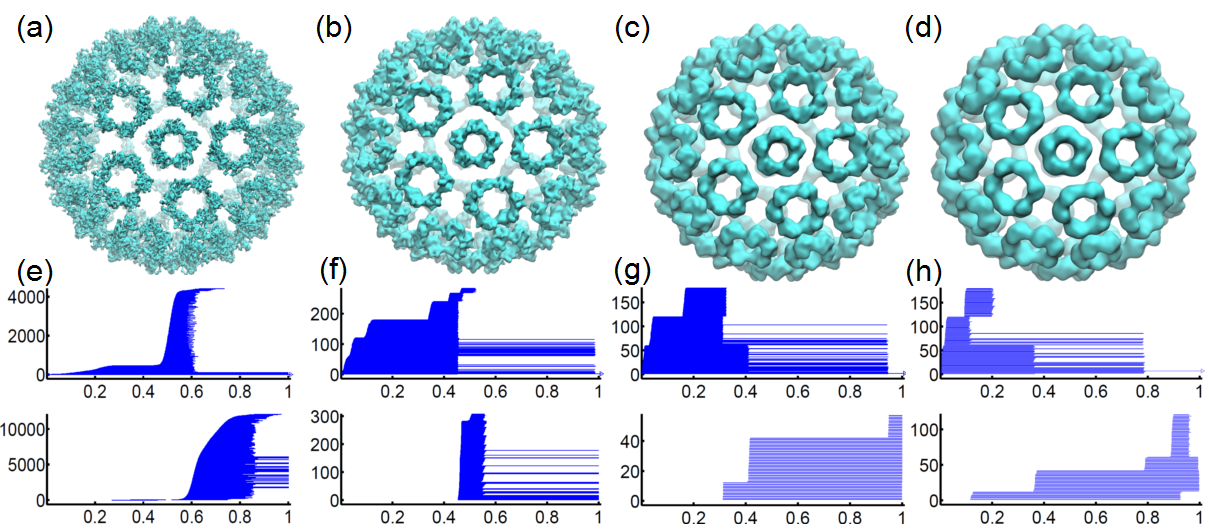}
\end{tabular}
\end{center}
\caption{ Multiresolution   analysis of virus capsid structure generated from protein 1DYL.
({\bf a})-({\bf d}) Isosurfaces of rigidity density profiles generated at four resolutions $\eta= 2.0, 4.0, 6.0$ and 10.0\AA;
({\bf e})-({\bf h}) The corresponding persistent barcodes for the above rigidity density profiles. Top and bottom panels are for $\beta_0$ and $\beta_1$ barcodes, respectively.  The horizontal axises denote the rescaled rigidity density value.  It can be seen from  ({\bf e}) and ({\bf f}) that, topological invariants generated from the structure details within the protein are captured. Small scale topological details  quickly disappear as  the resolution parameter increases. Instead global topological invariants that capture the loops within or between protein-protein complex appear. In ({\bf g}), at the beginning of the filtration, 12 long and 30 short $\beta_1$ bars can be identified, which indicates  12 pentagon rings and 30 hexagon rings in the multiprotein complex. Later, 19 more $\beta_1$ bars emerge, corresponding  to 20 loops formed between hexagon and pentagon protein  complexes, with 1 loop being left uncounted because of a sphere-like surface \cite{KLXia:2015a}. The 20 loops further evolves into 80 loops as demonstrated in ({\bf h}). For the reason stated above, one loop is not counted.  More interestingly, by examining  birth and death time during the filtration, when $\beta_1$ bars emerge and $\beta_0$ bars disappear in ({\bf g}) and ({\bf h}), one can find that the hexagonal protein complexes contributes the first 60 $\beta_0$ bars. Only after a short while during the filtration, the second 60 $\beta_0$ bars from the pentagonal protein complex appear. Finally, the third 60 $\beta_0$ bars come out. They are from the short bars in hexagonal protein complexes as  seen in Figure \ref{fig:PC_hexagon_pattern}({\bf f}).
}
\label{fig:1dyl_multiscale}
\end{figure}

In hexagon-shaped protein complex, most of the above patterns are observed except for a very unique topological signature in $\beta_0$ panels at resolutions of $\eta=6.0$ \AA.  Unlike the pentagon counterpart, in which five  $\beta_0$ bars come out and die simultaneously, the hexagon protein complex does not generate six uniform $\beta_0$ bars due to the dimerization. At $\eta= 6.0$\AA, there are only 4 $\beta_0$ bars, two appearing together at first and another pair emerging later, as demonstrated in Figure \ref{fig:PC_hexagon_pattern}({\bf f}).

\begin{figure}
\begin{center}
\begin{tabular}{c}
\includegraphics[width=0.9\textwidth]{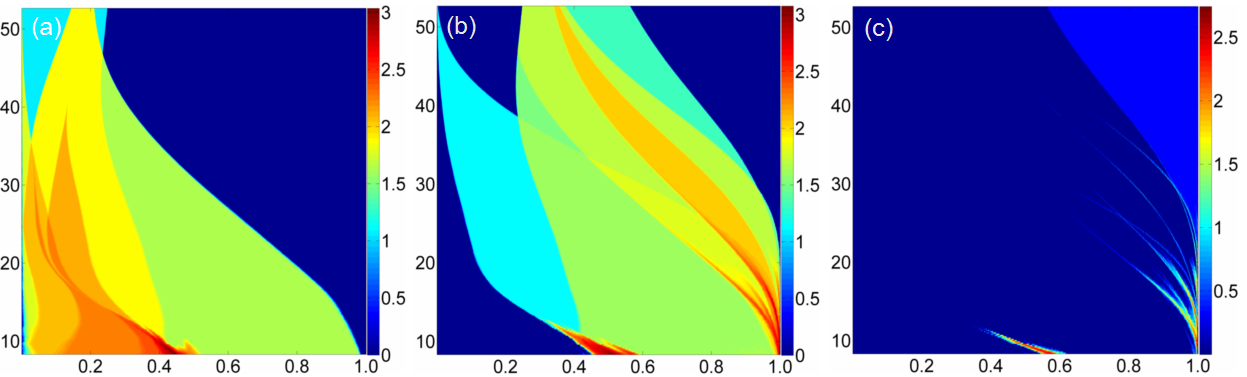}
\end{tabular}
\end{center}
\caption{ 2D persistent homology  analysis of virus capsid structure generated from protein 1DYL.
({\bf a})-({\bf c}) The logarithm plots of 2D persistent barcode numbers for $\beta_0$, $\beta_1$ and $\beta_2$, respectively. The horizontal axises denote the rescaled rigidity density value. Vertical axises are resolution $\eta$(\AA). It is worth noticing that four types rings in the $\beta_1$.
}
\label{fig:1dyl_multidimension}
\end{figure}

For the virus capsid structure, the topology at the multprotein scale reflects 12 pentagons and 30 hexagons. There are about 20 triangle circles formed between pentagon rings and hexagon rings. Each of these triangle circles can evolve into 4 smaller circles in different rigidity density isosurfaces when suitable resolutions are employed. All the above information can be derived from the careful observation of  $\beta_1$ panels in Figure \ref{fig:1dyl_multiscale}. Especially, in Figures \ref{fig:1dyl_multiscale}({\bf g})-({\bf h}), we can see that 12 $\beta_1$ bars emerge first during the density filtration. 
Additionally,  there are 30 long persisting bars in the $\beta_1$ panels due to 30 hexagonal rings. Furthermore,  there are 19 short bars, which are originated from 20 triangle circles, as the whole surface is connected and one $\beta_1$ bar is removed. Finally, the 60 more $\beta_1$ bars are added as each triangle transforms into 4 circles. It is interesting to analyze  $\beta_0$ barcodes as well. In Figures \ref{fig:1dyl_multiscale}({\bf g})-({\bf h}), there are roughly three sets of 60 $\beta_0$ bars appearing according to their generation time during the filtration process. From the above analysis, we know that 12 pentagon complexes contribute 60 identical $\beta_0$ bars.  Due to the dimerization,  30 hexagons complexes contribute two types of $\beta_0$ bars, each having 60 bars. Furthermore, from the birth and death time of  the first and second sets of $\beta_0$ bars, we can tell that the only the second set of 60 $\beta_0$ bars is due to the pentagon protein complexes, as 12 $\beta_1$ bars appear at exactly the same time. 

Figure \ref{fig:1dyl_multidimension} shows 2D persistence in the virus capsid structure. The $\beta_1$ information is particularly interesting. Clearly there are four types ring structures as analyzed above that coexist over a wide range of resolutions.

\subsection{Multiresolution topology based protein domain classification}

\begin{figure}
\begin{center}
\begin{tabular}{c}
\includegraphics[width=0.9\textwidth]{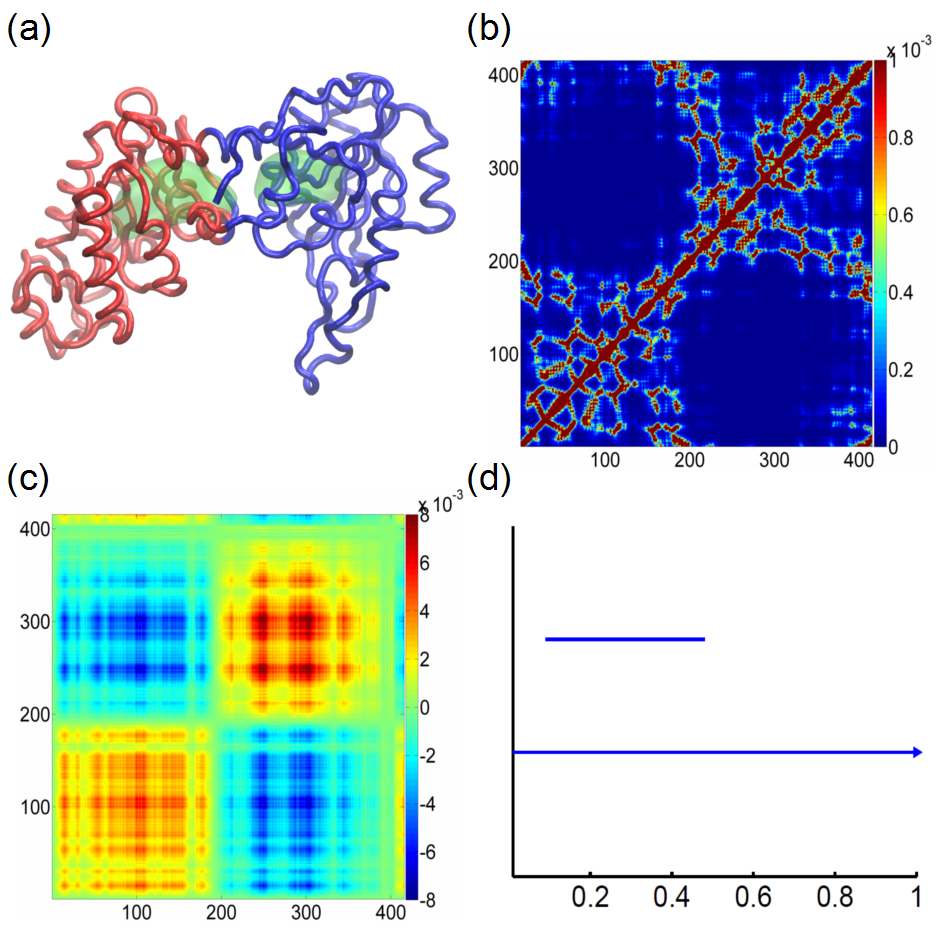}
\end{tabular}
\end{center}
\caption{ The domain partition of protein 3PGK. The coarse-grained C$_{\alpha}$ model is used. In our FRI correlation map, FRI matrix and rigidity density,  the Lorentz kernel is employed with $\upsilon=2$ and $\eta=5$ \AA.~
({\bf a}) The domain partition using the second smallest eigenvalue of our FRI matrix. The two domains are colored in red and blue, respectively. It should be noticed that two green surfaces within the protein are extracted from our rigidity density map, indicating two isolated components in topological analysis. ({\bf b}) The FRI correlation map constructed by $C_{ij}=\Phi( r_{ij};\eta)$.  ({\bf c}) FRI matrix constructed by the tensor product of the second eigenvector. ({\bf d}) The $\beta_0$ barcodes for the FRI rigidity density. It can be seen that using our multiresolution representation, we can identify two different domains in the protein.
}
\label{fig:Domain_3pgk}
\end{figure}

A protein domain is a relatively conserved part of a protein structure and has its own independent functions and structural shape. Protein domains serve as main building blocks for many large proteins and play an important role in protein design. For a given protein, identification and classification of protein domains is a crucial task. Gaussian network model (GNM) \cite{Bahar:1997}, FRI and graph theory  can be used for protein domain analysis.  In this work, we illustrate the use of proposed multiresolution persistent homology for protein domain classification, together with two other methods. To our knowledge, it is the first time that persistent homology is used for practical protein domain identification.

Figure \ref{fig:Domain_3pgk} illustrates the domain predictions by three methods. Protein 3PGK has two domains as shown in Figure \ref{fig:Domain_3pgk}(a). The FRI correlation map 
$\{C_{ij}\}_{i,j=1,\cdots,N}$ generated by $C_{ij}=\Phi( r_{ij};\eta)$ is shown in Figure \ref{fig:Domain_3pgk}(b). Clearly, first 200 residues form one domain and the rest residues belong to another one. Two domains are linked through an alpha helix. In  Figure \ref{fig:Domain_3pgk}(c), we plot the domain prediction from the second eigenvector of the FRI matrix
\begin{eqnarray}\label{eqn:MKirchhoff}
\Gamma_{ij}(\Phi )  = \begin{cases}\begin{array}{ll}
       - \Phi( r_{ij};\eta), &i\neq j  \\
        -\sum_{j, j\neq i}^N\Gamma_{ij}(\Phi),  & i=j
							\end{array}
       \end{cases}.
\end{eqnarray}
In this approach, we plot  matrix ${\bf u}_2{\bf u}^T_2$, where ${\bf u}_2$ is the second eigenvector of $\{\Gamma_{ij}(\Phi )\}_{i,j=1,\cdots,N}$  and $T$ denotes the transpose.  This matrix shows a clear separation of two domains as well. Finally, the $\beta_0$ panel of present persistent homology gives rise to two distinct bars, indicating the existence of two domains as well.

\section{Conclusion}

Persistent homology is a promising method for capturing the multiscale feature in data. However,  persistent homology does not create a multiscale representation on a data that has no multiscale trait. Instead, it applies a uniform resolution on any scale in the radius based filtration.  Since cross-scale filtration with a fine resolution is computationally unreachable, current persistent homology fails to work for large multiscale data, such as multiprotein complexes, which typically have atomic, residual, domain, protein and  protein-complex scales and may consist of millions of atoms in the data.  In this work, we introduce multiresolution persistent homology to overcome this difficulty. Our essentially idea is to choose appropriate resolution to match the scale of interest in topological analysis.  Therefore, in our approach, low resolution is applied to the analysis of large scale features whereas high resolution is reserved for small scale details. As a result, by tuning the resolution, we can focus the topological lens on the scale of interest in a big data set.


We construct the multiresolution persistent homology by extending flexibility-rigidity index  (FRI) originally proposed for biomolecular data to general data \cite{KLXia:2013d,Opron:2014}. The FRI method incorporates resolution-tunable kernel functions to measure the topological connectivity of a data set via a generalized distance and thus gives rise to a rigidity function for the underlying data. Such a rigidity function provides a matrix or volumetric density representation of the data set. Therefore, by an appropriate selection of the resolution, FRI based density filtration generates  resolution-matched persistent homology analysis at any specified scale.  Additionally, the present FRI method also provides a multiresolution geometric representation of the data set to match the scale of interest.

 We validate the proposed multiresolution topological method by a hexagonal fractal image which has a three-scale structure.   We show that by an appropriate choice of resolution, the proposed method is able to capture the topology at each of the three scales. We further illustrate the proposed multiresolution geometric and topological analysis by a few biomolecules. Multiscale persistent homology analysis is carried out by using a DNA molecule with all-atom, all-atom without hydrogen and  coarse-grained (nucleic acid) representations. The topological fingerprints generated these representation differ from each other, implying the potential complication and inconsistency in multiscale persistent analysis. In contrast to the multiscale persistent homology analysis, the proposed multiresolution persistent homology analysis is achieved based on the all-atom data. In this approach, resolution based continuous coarse-grained representation at any desirable scale can be constructed.   The utility of the proposed method is also investigated by using an RNA molecule and a virus complex consisting of 240 protein monomers, which is too large to be studied by the point cloud method. The desirable topological fingerprints of the virus multiprotein complex are revealed in our numerical experiments.  Finally, we apply the proposed multiresolution topological method to the protein domain identification. We show that by selecting the resolution to match the size of protein domains, the present method can effectively distinguish domains in a protein. 

We believe that proposed multiresolution persistent homology provides a general and practical approach for the topological simplification of big data in point cloud, pixel and voxel formats.  The present multiresolution approach can be directly applied to the geometric and topological analysis of general data sets, such as social networks, biological networks and graphs.

\section*{Acknowledgments}

This work was supported in part by NSF grants  DMS-1160352 and IIS-1302285,  NIH Grant R01GM-090208 and MSU  Center for Mathematical Molecular Biosciences initiative. The authors acknowledge the Mathematical Biosciences Institute for hosting valuable workshops.

\vspace{0.6cm}
%

\end{document}